# New g%AIC, g%AICc, g%BIC, and Power Divergence Fit Statistics Expose Mating between Modern Humans, Neanderthals and other Archaics

**Peter J. Waddell[1,2] and Xi Tan[2]**

pwaddell@purdue.edu
tan19@purdue.edu

[1]Department of Biological Sciences, Purdue University, West Lafayette, IN 47906, U.S.A.
[2]Department of Computer Science, Purdue University, West Lafayette, IN 47906, U.S.A

Weighted Least Squares trees (WLS), Multi-Dimensional Scaling (MDS), Balanced Minimum Evolution (BME), Neighbor Joining (NJ), and NeighborNets

The purpose of this article is to look at how information criteria, such as AIC and BIC, interact with the g%SD fit criterion derived in Waddell et al. (2007, 2010a). The g%SD criterion measures the fit of data to model based on a normalized weighted root mean square percentage deviation between the observed data and model estimates of the data, with g%SD = 0 being a perfectly fitting model. However, this criterion may not be adjusting for the number of parameters in the model comprehensively. Thus, its relationship to more traditional measures for maximizing useful information in a model, including AIC and BIC, are examined. This results in an extended set of fit criteria including g%AIC and g%BIC. Further, a broader range of asymptotically most powerful fit criteria of the power divergence family, which includes maximum likelihood (or minimum $G^2$) and minimum $X^2$ modeling as special cases, are used to replace the sum of squares fit criterion within the g%SD criterion. Results are illustrated with a set of genetic distances looking particularly at a range of Jewish populations, plus a genomic data set that looks at how Neanderthals and Denisovans are related to each other and modern humans. Evidence that *Homo erectus* may have left a significant fraction of its genome within the Denisovan is shown to persist with the new modeling criteria.

## 1 Introduction

Our purpose here is to look at how information criteria, such as AIC and BIC (described below), interact with the g%SD fit criterion derived in Waddell et al. (2007, 2010). The g%SD criterion measures the fit of data to model based on a geometric percentage standard deviation between the observed data and model estimates of the data, with 0% SD being a perfectly fitting model. However, it may not be adjusting for the number of parameters in the model comprehensively. We also turn our attention to a broader range of asymptotically most powerful fit criteria that includes maximum likelihood (or minimum $G^2$) and minimum $X^2$ modeling as special cases. Results are illustrated with a set of genetic distances looking particularly at a range of Jewish populations, and genomic data that looks at how Neanderthals and Denisovans are related to each other and modern humans.

The study of maximizing the useful information in a statistical model got a major boost from Akaike (1973, 1974) showing how the model with the expected minimum Kullback-Leibler (KL) divergence to new data might be inferred. This Akaike Information Criterion (AIC) has turned out to be of major benefit when the purpose of modeling is to predict new data, when the author is not too concerned about whether the set of parameters selected for the modeling is converging on a true set of parameters as more data is gathered, that is asymptotically with the amount of data. The AIC criterion also benefits from adjustments for bias, such as in AICc, when $k$ (the number of estimated parameters) is appreciable in comparison with the number of pieces of information, $N$; the exact adjustment does depend on the assumed error distribution of the data (e.g., Burnham and Anderson 2002). In contrast, the Bayesian Information Criterion (BIC, Schwartz 1978) is mostly concerned with a penalty function that will ensure exactly the true set of parameters generating the data will asymptotically be selected when the true model is accessible. This higher penalty means that the KL distance is not necessarily minimized, so some predictive power is lost. In some areas such as phylogenetics and historical population genetics, the major aim is not so much to predict new data, but to be confident that true historical features or facts, such a lines of descent, are accurately discovered. BIC has been shown to work surprisingly well compared to even more sophisticated Bayesian methods when it comes to accurately selecting a set of edges in a graph (e.g., Madigan and Raftery 1991).

Our purpose here is to show some relationships between g%SD, AIC and BIC and to produce hybrid measures of fit. We also extend our consideration to the power divergence family of fit (the $PD^{\lambda}$ statistic, Cressie and Read 1984), which includes measures such as maximum likelihood, plus minimum $X^2$, KL and Hellinger divergence, within it. We extend this family to more freely model errors as a function of the size of a piece of data, and find that the new statistic $PD^{a,\lambda}$ seems to yield interesting and potentially more accurate parameter estimates despite model inadequacies. That is, it appears a good candidate for a robust parameter estimation and model selection criterion.

Novel statistics and methods are illustrated with a set of genetic distance data looking at the origins of "Abraham's Children", including the Jews. A set of discrete genome-wide sequence site patterns, that includes Neanderthals and another archaic human (the Denisovan), as well as distantly related modern human genomes, are also fit. One of the interesting features of the latter data set is how exact modeling of the full site-pattern frequency spectrum is revealing surprising details of how archaic humans interbreed with modern humans and potentially, also with earlier species of archaic humans such as *Homo erectus* (Waddell et al. 2011).

## 2 Materials and Methods

The distance data used below are from table 1 of Atzmon et al. (2010), to eight decimal places of precision as communicated by Li Hao and as used by Waddell et al. (2010).

The genomic sequence data are those of Reich et al. (2010), based on alignments by Martin Kirsch, and communicated by Nick Patterson (as used in Waddell et al. 2011). These are reduced to 146,019 informative site patterns since the individual genomes are processed as pseudo-haploids (that is, in the individuals at the tips of each gene tree, only one state is scored) and because the singleton site patterns are most severely affected by sequencing error. Calculations for this data were performed in Excel using a spreadsheet developed in Waddell et al. (2011) to fit an infinite sties spectrum from a coalescent model to data. The main change herein is the use of a wide range of fit statistics, including in them a number of free parameters.

## 3 Results
### 3.1 The nature of the g%SD penalty compared to AIC, AICc and BIC

In Waddell et al. (2007, 2010a), we derive a fit measure that is monotonic with likelihood, that is the g%SD measure. This measure also has a simple geometric interpretation, that is, it is the average root weighted mean square percentage deviation using a geometric mean of the weights and distances for normalization. Let $n$ be the number of objects that distances are recorded amongst, $N$ is then the total number of distinct informative distances (the $d_{obs}$) of which there are typically $n(n-1)/2$ in a symmetric matrix with the diagonal equal to zero. These distances each have a weight, e.g. of the form $1/d_{obs}^P$ (where $P$ is a real number, assuming the distances are positive numbers). Recall that the g%SD of the data to the model with $k$ fitted parameters is equal to

$$g\%SD = \left(GM[w_{d_{obs}}]\right)^{0.5} \times \left(GM[d_{obs}]\right)^{-1} \times \left(\frac{1}{N-k}SS\right)^{0.5} \times 100\% \qquad (eq\ 1),$$

where $GM$ indicates a geometric mean and $SS$ is the weighted residual sum of squares, which includes the free parameter $P$ to model the form of the residual errors. The estimated log likelihood of the data, with respect to variance weighting factor $P$, is $\ln \hat{L}_P$. Recall equation S4 from Waddell et al. (2012, submitted), that is

$$\ln \hat{L}_p = \frac{n(n-1)}{2}\left(-\frac{1}{2}\ln[2\pi] - \frac{1}{2}\ln\left[\frac{g\%SD}{100}\right] - \ln[GM[\mathbf{d}_{obs}]] - \frac{1}{2} - \frac{1}{2}\ln\left[\frac{N-k}{N}\right]\right)$$
$$\sim \frac{n(n-1)}{2}\left(-1.41894 - \frac{1}{2}\ln\left[\frac{g\%SD}{100}\right] - \ln[GM[\mathbf{d}_{obs}]] - \frac{1}{2}\ln\left[\frac{N-k}{N}\right]\right) \qquad (eq\ 2)$$

Here it can be seen that minimizing g%SD for a given data set is the same as maximizing

$$\ln \hat{L}_P + \frac{n(n-1)}{2}\left(\frac{1}{2}\ln\left[\frac{N-k}{N}\right]\right) = \ln \hat{L}_P - \frac{N}{2}\ln\left[\frac{N}{N-k}\right],$$

which in turn is the same as minimizing

$$N\ln\left[\frac{N}{N-k}\right] - 2\ln \hat{L}_P \qquad (eq\ 3)$$

This last form includes a penalty term of the form $N\ln(N/(N-k))$ which is initially smaller than that used by the well-known AIC penalty of $2k$ (Akaike 1973). However, as $k \to N$ this penalty approaches and finally exceeds even the larger BIC penalty term of $k\ln[N]$ when $k = N-1$.

Figure 1 shows the total penalty terms as they would apply to a moderate sized data set with 17 tips and hence $N = 136$ unique distances, for each possible value of $k$ from 0 to $N$-1 (such as the data of Atzmon et al. (2010) as used in Waddell et al. (2010b)). For small $k$, the penalty of g%SD (= $N\ln(N/(N-k))$) is about ½ that of AIC (= $2k$). The AICc penalty term (= $2k/(N-k)$), a version of AIC corrected for small sample bias (Sugiura 1978), rises rapidly as $k$ approaches ½$N$. The g%SD penalty only exceeds that of AIC for relatively large $k$, as might rarely be encountered with a potentially massively parameterized model such as NeighborNet (Bryant and Moulton 2004, or other general graphs such as graphical models, Madigan and Raftery 1991, Waddell and Kishino 2000) or else a very small data matrix between just a few objects. The BIC penalty (Schwartz 1978), is a dimensionally stable or consistent criterion (that means that in the limit, it may select only the relevant parameters, nothing more and nothing less), starts off largest, but is overtaken at just over ½ $N$ by the AICc penalty.

Note, that our $N$ here is the number of input distances. $N$ could, for example, also be counted by measuring how many aligned nucleotides contributed to estimating the distances. However, this leaves many ambiguities, such as whether to count constant site patterns and whether to use the full dimension of the alignment matrix. It only makes full sense to preserve detailed information about $N$ in the variances and model selection if we are confident of the model that generated the nucleotide sites. For example, if each nucleotide site is strongly correlated with another, then the amount of information may be closer to ½ $N$ rather than to $N$. Such issues are sidestepped in the model evaluations herein where the likelihood is closer to a quasi-likelihood (Wedderburn 1974), based on the residual error we measure between the data and the model, not what we a priori imagine reality is like. Burnham and Anderson (2002) discuss QAIC and rederive AIC and AICc estimates. For simplicity, while some of our likelihood functions used herein are close to quasi-estimates, we retain the standard labels. Part of the reason for this is that there are different concepts of what QAIC can be. For some of our data we know that the inflation factor due to sampling effects is probably less than 2. However, when we model the total error structure, we are dealing with inflation factors of 10 to 30, which is well outside the range considered usual, and is moving more into the area of robust modeling than just quasi-likelihood inflation adjustment as considered in Burnham and Anderson (2002). Perhaps in future an appropriate term might be RAIC, etc. for RobustAIC.

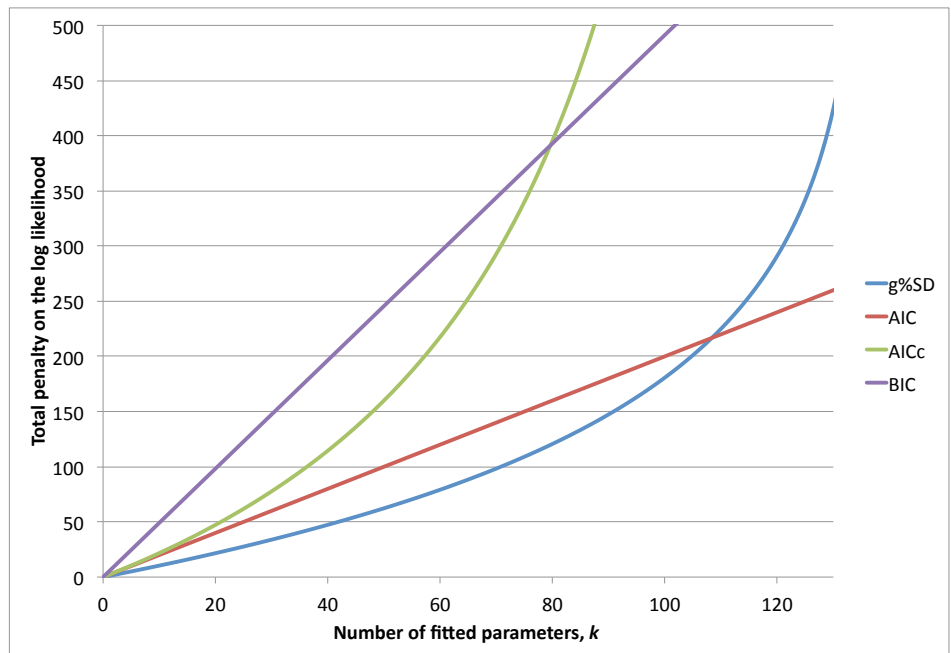

Figure 1. The relative penalties of the fit statistics g%SD, AIC, AICc and BIC for $N$=136 unique distances and various $k$. These penalties are traditional considered to apply in log likelihood space (e.g., Akaike 1973).

### 3.2 Deriving g%AIC, g%AICc and g%BIC

Recalling equation 1 and 2 above, in our case we have the relationship

$$\ln \hat{L}_p = \frac{n(n-1)}{2}\left(-\frac{1}{2}\ln[2\pi] - \frac{1}{2}\ln\left[\frac{g\%SD}{100}\right] - \ln[GM[\mathbf{d}_{obs}]] - \frac{1}{2} - \frac{1}{2}\ln\left[\frac{N-k}{N}\right]\right)$$
$$\sim \frac{n(n-1)}{2}\left(-1.41894 - \frac{1}{2}\ln\left[\frac{g\%SD}{100}\right] - \ln[GM[\mathbf{d}_{obs}]] - \frac{1}{2}\ln\left[\frac{N-k}{N}\right]\right) \quad \text{(eq 4)}$$

If we instead define the g%AIC to be

$$g\%AIC = \left(GM[w_{d_{obs}}]\right)^{0.5} \times \left(GM[d_{obs}]\right)^{-1} \times \left(\frac{1}{N}\exp\left[\frac{N+2k}{N}-1\right]SS\right)^{0.5} \times 100\%$$
$$= \left(GM[w_{d_{obs}}]\right)^{0.5} \times \left(GM[d_{obs}]\right)^{-1} \times \left(\frac{1}{N}\exp\left[\frac{2k}{N}\right]SS\right)^{0.5} \times 100\% \quad \text{(eq 5)}$$

then we have the relationship to the Akaike Information Criterion (AIC) of,

$$\frac{AIC}{N} = \left(2\ln\left[\frac{g\%AIC}{100}\right] + 2\ln(GM[\mathbf{d}_{obs}]) + 2\ln[2\pi] + 1\right) \quad \text{(eq 6)}$$

so,

$$AIC = \frac{n(n-1)}{2}\left(2\ln\left[\frac{g\%AIC}{100}\right] + 2\ln(GM[\mathbf{d}_{obs}]) + 2\ln[2\pi] + 1\right) \quad \text{(eq 7)}$$

Thus, mimimizing the g%AIC of equation 5 is the same as minimizing AIC. AIC is sometimes derived as a nearly unbiased estimator of leave out one observation (LOO) in some situations (Stone 1977). Thus, we might think of g%AIC as being close to the expected proportional error if we could expect to independently sample a replicate distance and compare it to the model distance.

The AIC penalty term in likelihood space, that is $\exp(2k/N)$ as in equation 5, can be approximated in various ways. A first order Taylor expansion yields $\exp(2k/N) \sim (N+2k)/N$, while a Euler approximation is $\exp(2k/N) \sim (N+k)/(N-k)$. A second order Taylor expansion is $\exp(2k/N) \sim (N^2+2Nk+k^2)/N^2$.

Similarly for AIC corrected for small sample sizes assuming normally distributed deviates (Sugiura 1978), that is g%AICc, we have

$$g\%AICc = \frac{\left(GM[w_{d_{obs}}]\right)^{0.5}}{\left(GM[d_{obs}]\right)} \times \left(\frac{1}{N}\exp\left[\frac{N+2k+2k(k+1)/(N-k)}{N}-1\right]SS\right)^{0.5} \times 100\%$$
$$= \frac{\left(GM[w_{d_{obs}}]\right)^{0.5}}{\left(GM[d_{obs}]\right)} \times \left(\frac{1}{N}\exp\left[\frac{N+2kN/(N-k)}{N}-1\right]SS\right)^{0.5} \times 100\% \quad \text{(eq 8)}$$
$$= \frac{\left(GM[w_{d_{obs}}]\right)^{0.5}}{\left(GM[d_{obs}]\right)} \times \left(\frac{1}{N}\exp\left[\frac{2k}{N-k}\right]SS\right)^{0.5} \times 100\%$$

so,

$$AICc = \frac{n(n-1)}{2}\left(2\ln\left[\frac{g\%AIC}{100}\right] + 2\ln(GM[\mathbf{d}_{obs}]) + 2\ln[2\pi] + 1\right) \quad \text{(eq 9)}$$

Note here we have used the form that is encountered when the observed and the expected data are constrained to sum to the same total. In the case of distances, the denominator $N-k-1$ is replaced with $N-k$. Thus, mimimizing the g%AICc is the same as minimizing AICc, which is an adjustment of AIC recommended whenever $N/k < 40$ (Burnham and Anderson 2002).

The g%AICc term $\exp(2k/(N-k))$ can be approximated with a first order Taylor expansion as $\exp(2k/(N-k)) \sim (N+k)/(N-k)$ or a Euler approximation of $N/(N-2k)$. The quite intuitive Euler approximation is particularly close for low values of $k$ to $N$, as we will see later in figure 2. Alternatively, and of potential use with constrained vectors of counts as we encounter later, the g%AICc term $\exp(2k/(N-k-1))$ can be approximated with a first order Taylor expansion as $\exp(2k/(N-k-1)) \sim (N+k-1)/(N-k-1)$ or a Euler approximation of $(N-1)/(N-2k-1)$. These approximate relationships show that AICc is rather like g%SD, except that the correction is double the penalty per fitted parameter. That is, g%SD uses a bias correction term of $1/(N-k)$ while the Intuitive Euler approximation of $1/N\exp(2k/(N-k))$ is $1$ $(N/(N-2k))/N = 1/(N-2k)$. This may be a gentle way of introducing AICc into simple regression equations when $\hat{\sigma}^2$ is estimated from the residuals.

Finally, for g%BIC, we have

$$g\%BIC = \left(GM[w_{d_{obs}}]\right)^{0.5} \times \left(GM[d_{obs}]\right)^{-1} \times \left(\frac{1}{N}\exp\left[\frac{(k-N)\ln[N]}{N}\right]SS\right)^{0.5} \times 100\%$$

(eq 10)

$$= \left(GM[w_{d_{obs}}]\right)^{0.5} \times \left(GM[d_{obs}]\right)^{-1} \times \left(N^{(k/N-1)} \times SS\right)^{0.5} \times 100\%$$

Where the last substitution occurs because $\exp[k\ln[N]/N]=N^{k/N}$. While minimizing g%BIC will also minimize BIC, it is less clear what the g%BIC is measuring. This is because the properties of BIC are not so clear in terms of cross validation or prediction. The BIC penalty guarantees dimensional consistency if the true model is in the set (and given certain assumptions). In terms of trees, this means that, asymptotically for large amounts of data under the model, the correct tree is selected, even if it is a non-binary tree (Waddell 1998).

Therefore, perhaps g%BIC can be thought of as the prediction accuracy expected for a model that is expected to be dimensionally consistent asymptotically. Note, the BIC penalty can be smaller than AICc in our QAIC-like context. In this situation the $N$ used does not grow as more data is used to estimate the distances, but $k$ can change while $\hat{\sigma}^2$ will hopefully keep decreasing. Looking at the form of equation 8, and the fact that that AICc out-penalizes BIC in figure 1, to restore dimensional stability in our context may require the penalty $\exp((k\ln[N])/(N-k))= N^{k/(N-k)}$, which we will call BICc. It is anticipated that g%BICc is giving us an estimate of the prediction accuracy of a model that is expected to contain a set of "real" parameters as $N$ gets larger due to adding in more objects and/or as $\hat{\sigma}^2$ hopefully goes to zero. Dimensional consistency as promised by BIC has a special appeal in population genetics and particularly phylogenetics (as well as the crossover area of, phylopop) for certain types of parameter. For example, for collections of organisms each from very distinct taxonomic groups, a very strong argument can often be made that their whole genomes have evolved according to a single unweighted rooted tree. Thus, the edges identified in the model are hoped to correspond to real historical facts. Given that a prime consideration of phylogenetics is to establish these facts, dimensional consistency is especially important and in practice will often qualitatively exceed in importance any special need to predict a new piece of data or data set. In contrast, parameters associated with the substitution process are often of secondary importance and we recognize that parameters will be artificial to some degree. Thus, in the first instance of establishing the tree they are "nuisance parameters" and it seems acceptable to use AIC to minimize their impact on the precision of the model.

### 3.3 Relative penalties of g% measures and a worked example

To visualize the various approximations mentioned above, consider again the case of $N=136$ as shown in figure 2 below. For AIC, the first order Taylor expansion is not very good, but both the Euler and the second order Taylor expansion of AIC are reasonable up to $k$~35 or $k$ ~25% of $N$. More specifically, the error on estimating the total penalty exceeds 1 for the Taylor_1 approximation at $k = 8$, while for Taylor_2 the corresponding $k$ is 26 and for Euler 29. In percentage terms, the error on Taylor_1 is always greater than 1% for any $k$ of 1 or greater, while that of Taylor_2 is less than 1% up to and including $k = 18$, while for the Euler approximation it is up to $k = 23$. Thus, conceptually, the Euler approximation to AIC will be pretty close when $N > 135$ and $k < 20\%$ of $N$.

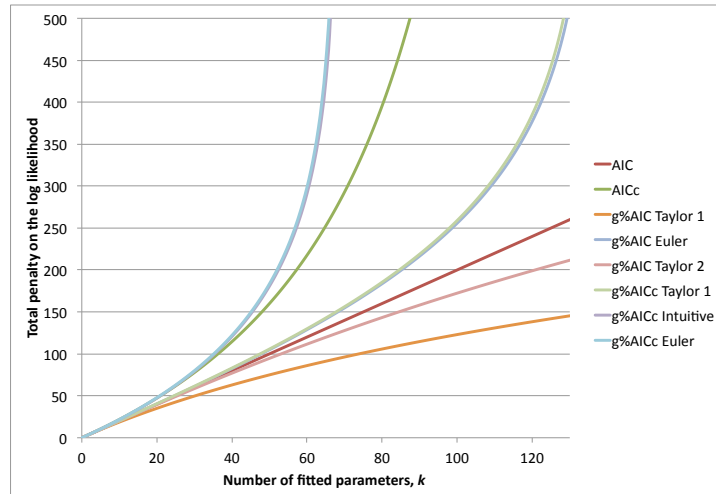

Figure 2. Approximations to the AIC and AICc penalties for $N = 136$. The penalty AICc Intuitive is $N/(N-2k)$, while AICc Euler is the Euler approximation to $\exp(2k/(N-k-1))$.

For AICc the first order Taylor expansion is clearly very poor, but the intuitively desirable Euler approximation of $1/(N-2k)$ for correcting the sum of squares term is good until around $k$~35. More specifically the error on the total penalty exceeds 1 for the Taylor 1 AICc expansion at $k = 7$, while for Euler it is a respectable 28. In percentage terms, the error on Taylor 1 is always greater than 1% for any $k$ of 2 or greater, while that of Euler is less than 1% up to and including $k = 25$. Thus, conceptually, the Intuitive Euler approximation for g%AICc of $1/(N-2k)$ will be pretty close to the true value when $N > 135$ and $k < 20\%$ of $N$. Thus this approximation is not only intuitively appealing but it should become very accurate for a slowing growing $k$ and larger $N$.

Table 1 shows the fit of g% measures to the data of Atzmon et al. (2010) as used in Waddell (2010b and updated for Systematic Biology, submitted 2012). The data are genetic distances between an African outgroup, ten Semitic populations (including seven widely scattered Jewish populations) and six European populations. The original data were single nucleotide polymorphisms, but here it is treated as thought the final form of the distances is all that is known of them (which is often the situation for types of data that are generated originally as distances). In table 1, for all but the g%BIC measures, the NeighborNet planar graph with edge lengths estimated by WSS with $P = 2$ is best. However, for the BIC terms, which seem most desirable in terms of selecting a graph with edges that have a factual basis, then the flexi-weighted least squares with polynomial weights seems best. Notice that the steps from g%SD to g%AIC to g%AICc are about equal in this example, although this relative scaling will change with the data as shown in figures 1 and 2. The precision (and predictive accuracy) of the best models by the g% criteria (excluding g%BIC) is between ~ 8 and 11%. This is only moderate, in that even single gene trees (e.g., those used in Waddell et al. 2009) tend to have g% values less than 4. The problem is not with the coalescent process per se, since distances on the "species" tree under a standard coalescent are still additive (as seen in Waddell et al. 2011). It must be something else and a strong contender is that there are many latent variables, which are substantial lines of descent (equals edges) in the true reticulate graphical model of descent.

For the two g%BIC models the penalty is markedly more severe. On the best models the fit is around 14 to 18% and for g%BICc a second tree models with the residuals of distances modeled by an exponential function, has overtaken the fit of the NeighborNet. For the very best model, the g%BICc is at 15.6%. It is speculated that what this percentage represents is a prediction (or expectation) of the precision of estimating a distance or data set using just those unidentified parameters we have in the current model that will remain in the highest posterior probability model (assuming the uniform priors that BIC uses). If that is correct, then the g%BICc measure may be of particular relevance when it comes to choosing an error structure to perform residual resampling (Waddell and Azad, 2009, Waddell et al. 2010a) when the primary purpose is to estimate the probability that a parameter is in the true model.

Table 1. Fit values for models of six forms of multi-dimensional scaling (MDS), NeighborNet (NN) and trees (the first five) applied to the Atzmon (2010) data), as used in Waddell (2010) which gives further details. Fit values include the g%AIC_E and g%AICc_E with the Euler approximation terms $(N+k)/(N-k)$ and $N/(N-2k)$, respectively,

| Model | WSS | P or P' | k | lnL | g% SD | g% AIC | g% AIC_E | g% AICc | g% AICc_E | g% BIC | g% BICc |
|---|---|---|---|---|---|---|---|---|---|---|---|
| NJ | 3.55E-5 | na | 31 | 550.2 | 30.2 | 33.3 | 33.5 | 35.6 | 36.0 | 46.4 | 54.8 |
| BME | 2.16E-5 | na | 31 | 602.6 | 20.5 | 22.7 | 22.8 | 24.2 | 24.5 | 31.6 | 37.3 |
| OLS+ | 5.43E-4 | 0 | 25 | 652.4 | 13.8 | 15.0 | 15.1 | 15.7 | 15.7 | 19.6 | 21.8 |
| fWLS-*P* | 0.11670 | 1.5 | 28 | 709.2 | 9.2 | 10.1 | 10.2 | 10.7 | 10.7 | **13.7** | **15.6** |
| fWLS-*P'* | 1.50E-4 | 21.0 | 28 | 695.7 | 10.2 | 11.2 | 11.2 | 11.8 | 11.9 | 15.1 | 17.2 |
| MDS-OLS+ | 0.00131 | 0 | 31 | 592.7 | 22.1 | 24.4 | 24.5 | 26.1 | 26.3 | 34.0 | 40.1 |
| MDS-*P* | 0.03814 | 0.9 | 32 | 616.4 | 18.6 | 20.6 | 20.7 | 22.2 | 22.4 | 29.1 | 34.7 |
| MDS-*P'* | 6.53E-4 | 13.2 | 32 | 612.0 | 19.2 | 21.3 | 21.4 | 22.9 | 23.1 | 30.0 | 35.8 |
| MDS3-OLS+ | 9.31E-4 | 0 | 47 | 615.7 | 20.3 | 23.2 | 23.5 | 27.8 | 29.5 | 38.3 | 60.0 |
| MDS3-*P* | 0.03229 | 1.0 | 48 | 655.9 | 15.2 | 17.4 | 17.6 | 21.0 | 22.5 | 29.0 | 46.6 |
| MDS3-*P'* | 2.91E-4 | 19.7 | 48 | 653.2 | 15.5 | 17.7 | 18.0 | 21.4 | 22.9 | 29.6 | 47.5 |
| NN-*P* | 0.69422 | 2.0 | 38 | 728.6 | **8.4** | **9.4** | **9.5** | **10.5** | **10.8** | 14.2 | 18.5 |

## 3.4 g%$X^2$, g%AIC, g%AICc and g%BIC for vectors

While distances are sometimes the fundamental form of data, we often work with a vector of values and attempt to fit it using maximum likelihood (= minimum $G^2$) or minimum $X^2$. We will start by considering something close to the latter, that is, the inverse or modified Neyman $X^2$, that is $iX^2$ (Neyman 1949). Note, since the observed and expected vectors are often normalized to sum to the same amount, that is a step in the analysis. It is important to distinguish whether this is considered to be losing one piece of information or if it is one more efficiently estimated parameter. That is because criteria such as AICc or BIC clearly distinguish $k$ from $N$ (e.g., $2k/(N-1) \neq 2(k+1)/N$). It seems appropriate to consider a piece of information has been lost, or at least that this scale parameter has not been efficiently estimated, since numerical examples showing a superior fit can be obtained with a freely estimated scale parameter matching the expected counts to the observed counts without the need for both vectors to sum to the same value. In the latter case we would certainly consider this to be an additional parameter.

The g%$iX^2$ measure can be derived as being equal to

$$g\%X^2 = (GM[w_i])^{0.5} \times (GM[\hat{\mathbf{s}}_{obs}])^{-1} \times \left( \frac{1}{N-k} \sum_{i=1}^{N} \frac{(\hat{\mathbf{s}}_{obs_i} - \hat{\mathbf{s}}_{exp_i})^2}{w_i} \right)^{0.5} \times 100\%$$

(eq 11)

$$= (GM[\hat{\mathbf{s}}_{obs}])^{0.5} \times (GM[\hat{\mathbf{s}}_{obs}])^{-1} \times \left( \frac{1}{N-k} \sum_{i=1}^{N} \frac{(\hat{\mathbf{s}}_{obs_i} - \hat{\mathbf{s}}_{exp_i})^2}{\hat{\mathbf{s}}_{obs_i}} \right)^{0.5} \times 100\%$$

by following the same basic working as in Waddell et al. (2007) in deriving the g%SD for distance data. That is, the g%$iX^2$ is monotonic with a likelihood (in this case a $X^2$) fit of the data to a model. Note that here the geometric mean of the $\hat{\mathbf{s}}_{obs}$ vector is being used to normalize the vector. Another way to normalize would be to require the vector $\hat{\mathbf{s}}_{obs}$ to sum to one before operating on it. That is an incomplete normalization for our purpose, since it does not take into account how many elements the vector has nor the essentially multiplicative nature of the terms. By normalizing by the geometric mean of the elements of the vector we are measuring the geometric percent standard deviation of the elements of the vector (subject to a set of weights).

Alternatively, where the weights are equal to $\hat{\mathbf{s}}_{exp_i}$, as in the usual $X^2$, we have

$$g\%X^2 = (GM[\hat{\mathbf{s}}_{exp}])^{0.5} \times (GM[\hat{\mathbf{s}}_{obs}])^{-1} \times \left( \frac{1}{N-k} \sum_{i=1}^{N} \frac{(\hat{\mathbf{s}}_{obs_i} - \hat{\mathbf{s}}_{exp_i})^2}{\hat{\mathbf{s}}_{exp_i}} \right)^{0.5} \times 100\%$$

This fit criterion was originally derived by Pearson (1900) and is closely linked to the likelihood of the data. As mentioned earlier, the form here follows that of the weighted sum of squares likelihood where the variance and a free parameter $P$, determining how the variance grows with the expected value, with are determined from the model residuals. With quasi-likelihood it is useful to separate two components: (1) A scale factor that deals with overdispersion (such as estimating the variance from the observed residuals). (2) The desire to "learn" or fit, a feasible error distribution causing the residuals to increase realism and robustness (something very important in situations were the errors are not at all well predicted by a fully parameterized likelihood model). Thus, it can also be considered that we are using here is a type of quasi-likelihood.

By replacing the normalization/penalty term of $1/(N-k)$ in with $1/N\exp(2k/N) \sim (N+k)/(N(N-k))$ we have the g%AIC version. Similarly, replacing $1/(N-k)$ with $1/N\exp(2k/(N-k-1)) \sim 1/(N-2k)$ we have the g%AICc version. And, finally, replacing $1/(N-k)$ with $N^{k/N-1}$ or $N^{k/(N-k)-1}$ we have the g%BIC or g%BICc version. In practice, or at least in thought, it can be useful to keep the $1/N$ term (which specifies the likelihood) separate from the penalty term, e.g. $\exp(2k/(N-k))$ which starts at 1 and increases. This later inflationary term is the effect upon the precision of the model in measuring expected amounts of divergence expected under different assumptions.

### 3.5 Generalizing g% for vectors

When we do not accept the model and are particularly concerned about overdispersion, we may instead estimate the variance via the data to model residuals, rather than from a priori assumptions. The quasi-likelihood framework can be extended further by not only allowing the weights to simply be a scalar of $\hat{\mathbf{s}}_{exp_i}$ but also a general function of $\hat{\mathbf{s}}_{exp_i}$. Thus, we can arrive at a more general g%f($\hat{\mathbf{s}}_{exp}$) as was done with distances in Waddell et al. (2010a). For example, for general polynomial weights we have

$$g\%[\hat{\mathbf{s}}_{exp}]^P = (GM[w_i])^{0.5} \times \left(GM[\hat{\mathbf{s}}_{obs}]\right)^{-1} \times \left(\frac{1}{N-k}\sum_{i=1}^{N}\frac{(\hat{\mathbf{s}}_{obs_i} - \hat{\mathbf{s}}_{exp_i})^2}{w_i}\right)^{0.5} \times 100\%$$

$$= \frac{(GM[\hat{\mathbf{s}}_{exp}])^{P/2}}{\left(GM[\hat{\mathbf{s}}_{obs}]\right)} \times \left(\frac{1}{N-k}\sum_{i=1}^{N}\frac{(\hat{\mathbf{s}}_{obs_i} - \hat{\mathbf{s}}_{exp_i})^2}{\hat{\mathbf{s}}_{exp_i}^P}\right)^{0.5} \times 100\%$$

(eq 12)

We could alternatively replace the sum of squares term with the Power Divergence Statistic, $PD^\lambda$, with a single free parameter $\lambda$ (sometimes called a Cressie-Read Statistic, after Cressie and Read 1984). The limiting distribution of this statistic under a multinomial model as the sample size, $x$, goes to infinity is $\chi^2_{d.f.}$, where with d.f. = $N - 1$. At $\lambda = 1$ the $PD^\lambda$ statistic reduces to the $X^2$ statistic. At $\lambda = -2$ we have the inverse or $iX^2$ statistic (sometimes called the Neyman (1949) modified $X^2$ statistic) discussed above. Other special cases of the $PD^\lambda$ statistic include $\lambda = 0$ for the log likelihood ratio statistic, $G^2$ (defined in the limit as $\lambda$ goes to zero), $\lambda = -1$ for the Kullback-Leibler statistic (also defined in the limit as $\lambda$ goes to -1), and $\lambda = -1/2$ for four times the Matusita distance (or eight times the squared Hellinger distance) = Freeman-Tukey $F^2$ statistic.

It is worth noting that while things like $X^2$, $G^2$ or $KL$ may be called a distance, they are closer to a distance squared. By taking square roots, the g% measures are closer to true metric distances. Further, $PD^\lambda$ are asymetric distances, sometimes called directed divergences. That is, $PD^\lambda_{ij}$ may not equal $PD^\lambda_{ji}$, except at value $\lambda = -1/2$, although they can be symmetrized in a fairly natural way by adding together the values obtained at $\lambda - 1/2$ and $-\lambda - 1/2$. This blended weight approach may give more sensitivity to both inliers and outliers (Basu et al. 2002), although as we will see later it is this directional sensitivity that can be exploited when data does not fit the model. This asymmetry is not such a problem for g%SD measures as all these are effectively measured relative to a perfect model (not just a saturated multinomial model with sampling error, although the expected value of this in the g% scale can be estimated from the degrees of freedom, with boundaries if need be).

With sparse data there is the issue that observed entries may go to zero and l'Hopital's rule only suggests setting these terms to zero when $\lambda > -1$, (although in practice any cell with the observed value equal to zero may be set to zero). Importantly, for any member of the family that is defined when the observed is equal to zero, the contribution of that cell to the total fit is equal to zero, so it need not be calculated. This is a considerable advantage with exponentially sparse data, as encountered in phylogenetics (e.g., Waddell et al. 2009). None of the 119 cells in our data have counts close to zero (all >> 5).

The form of $PD^\lambda$ that we use is defined as

$$PD^\lambda(\hat{\mathbf{s}}_{obs}, \hat{\mathbf{s}}_{exp}) = \sum_{i=1}^{N}\frac{2}{(\lambda+1)}\left\{\frac{\hat{\mathbf{s}}_{obs_i}}{\lambda}\left[\left(\frac{\hat{\mathbf{s}}_{obs_i}}{\hat{\mathbf{s}}_{exp_i}}\right)^\lambda - 1\right] - (\hat{\mathbf{s}}_{obs_i} - \hat{\mathbf{s}}_{exp_i})\right\}$$

(eq 13)

Is one that we particularly like. The essential quantity in equation 13 remains that of a ratio of observed to expected, which is effectively the "average" thing we want to minimize. It is standardized by $N$, $k$ and the scale of the data in the g% statistics. This quantity is intimately related to Renyi's (1961) alpha class generalized measure of entropy (= uncertainty or disorder), while its relationship to a range of robust methods is considered in Bera and Bilias (2002). Note, this form of $PD^\lambda$ is slightly different to that of equation 6.12 of Read and Cressie (1988) and emphasizes the subtraction of the residual and its true scale per cell. For visualizing the fit of vectors with $X^2$, a traditional advantage has been that each cell is positive and sums up give the total $X^2$ statistic. Also, in the form of eq 13 or 6.12 of Read and Cressie (1988), each cell or term of the summation has a marginal distribution that converges to a chi-square distribution with degrees of freedom equal to 1. Thus, the fit of each cell to the overall fit may be readily inferred. If the CR statistic is used as above, then all terms are non-negative and become equal to zero only if the observed and the expected of that cell are equal. The $PD^\lambda$ statistic also has a distribution, asymptotically under the multinomial with fairly standard assumptions, of a chi-square with d.f. equal to $N - k$ in fairly simple cases of efficient estimators and no boundaries in the parameter space. Another very useful property of the family is that when cells are grouped, then $PD^\lambda$ must decrease (Rathie and Kannappan 1972). This property was used in Waddell (1995) to provide tight bounds for branch and bound of the maximum likelihood of DNA sequences evolved on trees. It is equally applicable to the multi-tree coalescent model as used herein, since it is a property of the form of the data, not the model. It comes from a theorem in Rathie and Kannappan (1972), which basically says that when two independent cells are merged, the total information and hence the $G^2$ statistic, can only stay equal or decrease. Equation 13 has a form that is approximately

$$PD^\lambda \sim \sum_{i=1}^{N} \frac{\left(\hat{s}^\lambda_{obs_i} - \hat{s}^\lambda_{exp_i}\right)^2}{\lambda^2 \hat{s}^{2\lambda-1}_{exp_i}} \quad (eq\ 14)$$

Note, when evaluating the fit with $PD^\lambda$, the parameter estimates should be optimized with respect to each power of $\lambda$ considered. Thus, when $\lambda = 0$, a maximum likelihood estimator is used, and at $\lambda = 1$, a minimum $X^2$ estimate is made. Note, all estimators in this class are equally efficient asymptotically as data fitting the model becomes large. Further, it is unclear that $\lambda = 0$ (likelihood) always has optimal or desirable properties compared to other values of when the data deviate from model expectations (Read and Cressie 1988). However, $\lambda = 0$ does have the property of being Bahadur efficient in some cases. In contrast, $\lambda = 1$ is Pitman efficient under certain assumptions of sparseness. By contrast, $\lambda = -1$ or the KL statistic encounters the problem of being undefined when some cells are zero. Further, Read and Cressie (1988: 80) tend to suggest limiting useful values of $\lambda$ to between -5 and +5, although as we shall see below, for real data this range may be too narrow and extending the range until a minimum is found is our preferred approach.

Another important property of $PD^\lambda$ is that of partitioning the goodness of fit of the statistic into two components, that is the deviation of the adopted model expected values from the true model and the deviation (e.g., due to sampling) of the observed frequencies from the expected model frequencies. That is, $PD^\lambda(\mathbf{s}_{true} : \hat{\mathbf{s}}_{obs}) = PD^\lambda(\mathbf{s}_{true} : \hat{\mathbf{s}}_{exp}) + PD^\lambda(\hat{\mathbf{s}}_{exp} : \hat{\mathbf{s}}_{obs})$ (Read and Cressie 1988, p35). Thus, to estimate the total distance of the data from the true model requires knowing about the sampling variance and bias of the last two components. In the case of genomics, the latter can be estimated by the deviations between chromosomes, which are truly independent objects, as used in Waddell et al. (2011). As already mentioned, this quantity is about 155, indicated that since for our data $PD^\lambda$ is nearly always >1000, the vast majority of the problem is with the model being used.

Note, in our experience, our main concerns in modeling data are slightly different to those most closely addressed in Read and Cressie (1988). There, a prime concern was an a priori value of $\lambda$ that has maximum power to detect deviations from the model, particularly in the case of multinomial data. It turns out that $\lambda$ near 2/3 is near optimal for a fixed $\lambda$, probably due to the boundary induced by positive only counts causing a bias towards Poisson outliers (rather than inliers). In our experience, if the data does not fit the model well, it at least as desirable to make robust estimates, as it is to know that the data do not fit the model. In that case the question becomes which value of $\lambda$ offers the best prospect of robustness, that is, giving parameter estimates that are either relatively insensitive to observed cell values that deviate strongly away from the model expectations and hence exert undesirable leverage or close to their true values under adequate models.

It is also useful to note that the extension of $PD^\lambda$ statistics for use in AIC or BIC-like model selection is natural and is explored in Cressie (1996). There they are called PIC, although perhaps more memorably they could be known as PDIC, that is, Power Divergence Information Criterion. Later we show worked examples of the statistics in this section in section 3.8.

### 3.6 Very general g% fit statistics

In a probabilistic sense, all the results above treat the pieces of information as effectively independent, thus the fit is a sum of discrete terms. If the pieces (or patterns) of information are not independent and we have knowledge of how pairs of pieces of information co-vary, then this additional information can be represented in a variance-covariance matrix $\mathbf{V}$. Note that in the previous examples above it is effectively assumed that $\mathbf{V}$ is a diagonal matrix, i.e., $\mathbf{W}$. Thus with $\mathbf{W}$ the normalization term is the geometric mean of the weights, which is also the normalized determinant of $\mathbf{W}$ (that is, the product of the diagonal to the power $1/N$, where the last term is also the dimension of $\mathbf{W}$). If instead $\mathbf{V}$ has non-zero off diagonal entries, then the appropriate least squares measure of deviation becomes the minimum Mahalanobis or generalized least squares (GLS) distance. The iterated GLS criterion is also a maximum likelihood estimator under the assumption of multivariate normal errors, which is the maximum entropy (minimally informative) distribution in many situations. It may also be useful to transform the observed data into orthogonal quantities before measuring its geometric mean in order to perform normalization on the GLS distance. This is achieved by pre-multiplying by the Upper Triangular Cholesky matrix of $\mathbf{C}^{-1}$, (where $\mathbf{C}$ is the correlation matrix derived from $\mathbf{V}$) let's call it $\mathbf{U}_C^{-1}$. Then

$$g\%GLS = (\det[\mathbf{V}])^{1/(2N)} \times \left(GM[\mathbf{U}_c^{-1}\hat{\mathbf{s}}_{obs}]\right)^{-1} \times \left(\frac{1}{N-k}(\hat{\mathbf{s}}_{obs} - \hat{\mathbf{s}}_{exp})^t \mathbf{V}^{-1}(\hat{\mathbf{s}}_{obs} - \hat{\mathbf{s}}_{exp})\right)^{0.5} \times 100\%$$

(eq 15)

To apply this equation, there needs to be a transformation such as $w_{ij} = d^P_{obs_{ij}}$, to transform the variances and to allow the covariances to increase or decrease appropriately with the variances.

### 3.7 Fit functions combining power divergence with quasi-like variances

In the power divergence family, the main effect is to penalize positive residuals (outliers) more or less relative to negative residuals (inliers) (with a balance coming with $\lambda = 0$ or the $G^2$ fit). In contrast, the polynomial weights family of least squares fit used in the first sections works by applying a function with parameter $P$ to either the observed or expected data and tuning $P$ to maximize the likelihood and minimize the weighted squared residuals. It is tempting to combine both effects together and aim to minimize the resulting statistic. One form that suggests itself is

$$PD^{\alpha\lambda} = GeoMean[\hat{\mathbf{s}}_{exp}^{\alpha-1}] \sum_{i=1}^{N}\left[\left\{\frac{2}{\lambda(\lambda+1)}\hat{\mathbf{s}}_{obs_i}\left(\left(\frac{\hat{\mathbf{s}}_{obs_i}}{\hat{\mathbf{s}}_{exp_i}}\right)^\lambda - 1\right) - \frac{2}{(\lambda+1)}(\hat{\mathbf{s}}_{obs_i} - \hat{\mathbf{s}}_{exp_i})\right\}\frac{1}{\hat{\mathbf{s}}_{exp_i}^{\alpha-1}}\right] \quad \text{(eq 16)}$$

where parameter $\alpha$ behaves like power $P$ used earlier. Leading to the statistic

$$g\%xPD^{\alpha\lambda} = \frac{\left(GeoMean[\hat{\mathbf{s}}_{exp}]\right)^{\alpha/2}}{GeoMean[\hat{\mathbf{s}}_{obs}]}\left(x\frac{PD^{\alpha\lambda}}{N}\right)^{0.5} \times 100\%$$

where $x$ is $1/N$ multiplied by the inflation factor used to account for estimating parameters from the data for g%SD, g%AIC, g%BIC, etc. (note, for g%SD the inflation term is $N/(N-k)$, so $1/N \times N/(N-k) = 1/(N-k)$ ).

## 3.8 An illustrated example using archaic *Homo* genomes

It is useful to illustrate the fit statistics for discrete count data with the fit of the "Neanderthal" genomic data of Reich et al. (2010) to the single species tree hierarchical coalescent model developed in Waddell et al. (2011). The actual model is a tree of the form ((D,N):g5,(S,(Y,(F,(H,P):g1,):g2,):g3,)g4),) where the five coalescent parameters g1-g5 measured in the number of generations divided by the effective population size of the genes (which for autosomes is twice the effective population size of males and females since each carries two copies). The tips of the tree are genomes respectively, Denisova, Neanderthal, San, Yoruba, French, Han and Papuan. The five coalescent parameters are bounded to be non-negative, but for this data/model combination all entries are substantially away from the nearest boundary anyhow. As in Waddell et al. (2011) we condition on fitting just the informative site patterns, as the uninformative patterns are heavily influenced by sequencing error, and we do this by making the expected patterns sum to the observed informative site patterns. Here we will use revised spreadsheet calculations correcting some earlier computational errors.

For this Neanderthal data the geometric mean of the observed site patterns is 740.3. The best fit obtained minimizing $iX^2$ is, a value of 2693.40, which translates to a g%$iX^2$ (eq 11) of 17.94. That is, $740.3^{-0.5} \times (1/(119-1-5) \times 2693.40)^{0.5} \times 100\%)$. Minimizing g%$iX^2$, in this case, is exactly the same as minimizing $iX^2$. Minimizing $X^2$ yields 3150.86 (or a g%$X^2$ of 20.00, with the geometric mean of the expected values being 785.97), while minimizing g%$X^2$ directly, yields a slightly different value of 19.96 (with $X^2$ = 3161.98). Later, we will explain why the $iX^2$ statistic fits better than $X^2$ to this data.

If the Neanderthal data was i.i.d. and came from the model, then the $X^2$ or $iX^2$ statistic should have an expected value equal to the residual degrees of freedom or 119-1-5 = 113. Thus, the g%$iX^2$ would then be about $740.3^{-0.5} \times (1/(113) \times 113)^{0.5} \times 100\%$ = 3.68. The "effective" size of the real data, compared to the "model" data (Waddell and Azad 2009), as utilized by this model, is a ratio of ~ $(3.68/17.94)^2$ = (115/2693.40) = 0.0427, only 4.27%. Thus, there is considerably more useful information to be extracted from the data in theory, assuming that the original units of aligned nucleotides are not massively positively correlated.

The corresponding g%AIC-$iX^2$, g%AICc-$iX^2$, g%BIC-$iX^2$ and g%BICc-$iX^2$ values for the Neanderthal data be 19.85, 19.89, 21.05 and 21.14, respectively. In contrast, if the data were i.i.d. and came from the model, the values should be about 3.75, 3.76, 3.98, and 4.00, respectively. However, by comparing independent chromosome vectors, we recognize the possibility of overdispersion by a factor of about 155/(118-(118/23)) (figure 4, Waddell et al. 2011, with correction for the degrees of freedom of the global mean values each chromosome is compared to). This is the variance suggested by the most highly parameterized model we have at our disposal (actually, a saturated multinomial model when all autosomal chromosomes are considered to come from the same distribution). It is therefore sometimes recommended to use this value as a baseline for comparison (e.g., Burnham and Anderson 2002). In that case, the best g%AIC-$iX^2$, g%AICc-$iX^2$, g%BIC-$iX^2$ and g%BICc-$iX^2$ scores that should be expected are the previous values inflated by a factor of $(155/(118-(118/23)))^{0.5}$ ~ 1.173, yielding 4.50, 4.51, 4.77 and 4.79, respectively.

Figure 3 shows the relative performance of optimal fitting with $PD^\lambda$ and $PD^{\omega\lambda}$. For $PD^\lambda$ the fit curve is smooth with a minimum at $\lambda$ = -7.2. This is a surprisingly low value, given that nearly all the traditional power divergence fit statistics fall in the range of -2 to +1. This may be interpreted as the data having a major bias towards outliers rather then inliers. The corresponding g% statistics appear with the expected increasing magnitude, for small $k$ (here 5) compared to $N$ (here 118), of g%SD, g%AIC, g%AICc, g%BIC, and g%BICc. At this scale the most marked difference of these is between the BIC based statistics and the rest at about 1% out of a minimum of around 16%. Thus, the predictive power of AIC and BIC selected models seems very similar here.

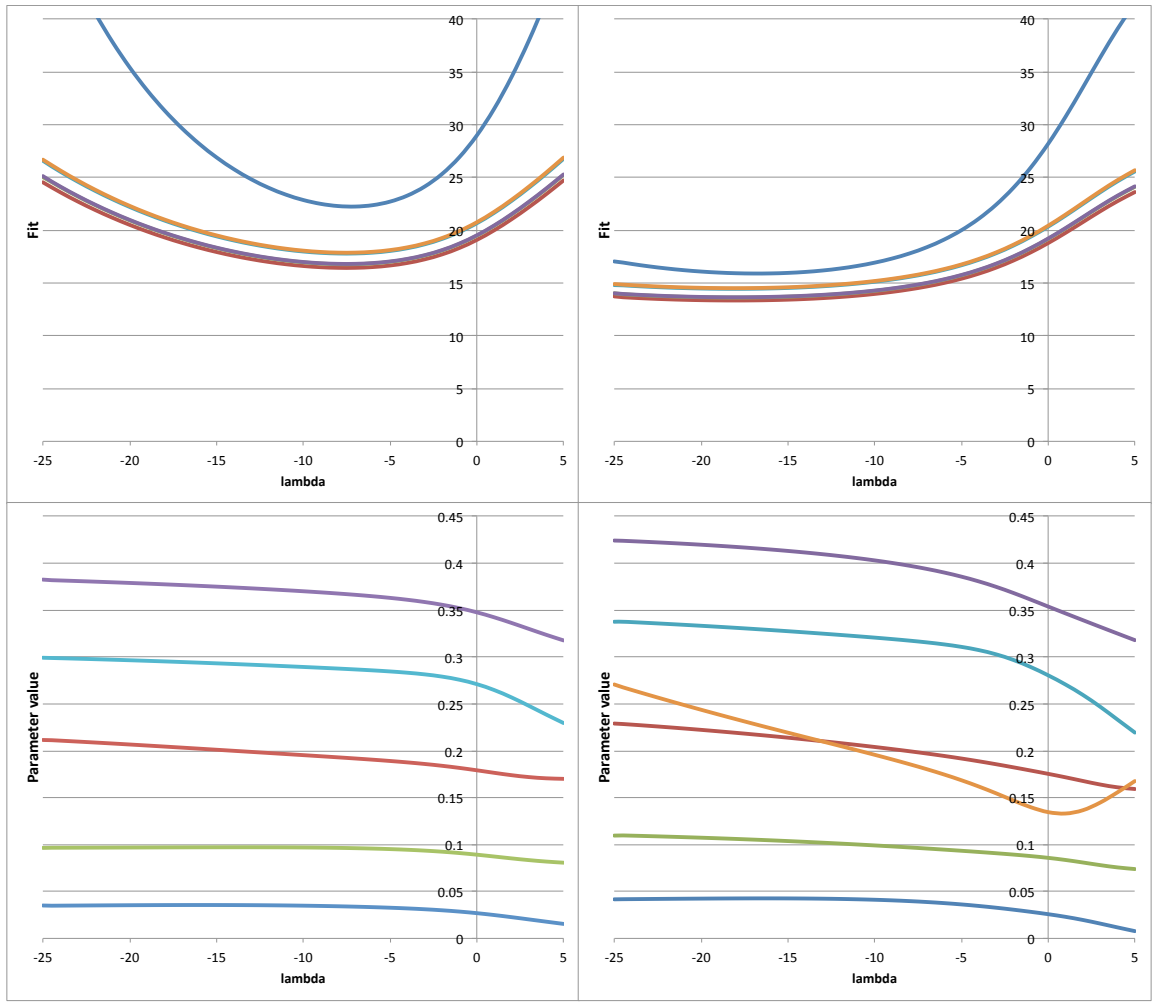

Figure 3. Fit of models and parameter estimates with $PD^\lambda$ and $PD^{\alpha\lambda}$. The upper left figure shows the optimal fit achieved with the $PD^\lambda$ statistic for various values of $\lambda$ (lambda). The upper blue line is the optimized $PD^\lambda$ statistic itself (divided by 100 for plotting), while the five colored lines in decreasing order are g%BICc_$PD^\lambda$, g%BIC_$PD^\lambda$, g%AICc_$PD^\lambda$, g%AIC_$PD^\lambda$ and g%SD_$PD^\lambda$. The upper left plots shows are the corresponding statistics when optimizing $PD^{\alpha\lambda}$. The lower left plot shows the parameter estimates for $PD^\lambda$ in order from smallest to largest being, g1, g3, g2, g5, g4. The lower right plot shows the corresponding parameter values obtained by optimizing $PD^{\alpha\lambda}$; also shown in orange is the optimal value of parameter α (divided by 10 for plotting).

The actual parameter estimates grow as $\lambda$ becomes more negative, but have flattened out considerably by the optimal $\lambda$ at around -7.2. It is hard to know what to compare these estimates to, however there are some contenders based on methods yielding better fits. One of these, on page 22 of Waddell et al. (2011), is to minimize the percent deviation (not the root mean square

Waddell and Tan (2012), *g%AIC, Power Divergences Statistics and Neanderthal Mating.*    Page 14

percentage) weighted by the expected size of the pattern. Using the revised site pattern calculations, this fit criterion yields a minimum %SD_ABS of 12.9 and parameter estimates g1-g5 of 0.0295, 0.1798, 0.0964, 0.3927 and 0.2970, respectively. These parameter estimates are all larger than those obtained with $G^2$ (ML) or $\lambda = 0$ (table 2), particularly those for g4 and g5.

Table 2. Parameter estimates for the Neanderthal data using methods: M1, Approximate Bayesian Calculation (ABC) quadratic regression. M2, ABC quadratic regression on each marginal distribution separately. M3, fitting with the $PD^\lambda$ statistic at the optimal $\lambda$ value of -7.2. M4, fitting with the $PD^{\omega\lambda}$ statistic at the optimal $\lambda$ of -16.8 and $\alpha$ = 2.3. Then the difference squared of each parameter pair for each model. For comparison the ML estimates are 0.03, 0.18, 0.09, 0.35 and 0.27.

| Param. | M1 | M2 | M3 | M4 | $\Delta^2$ | M1:M2 | M1:M3 | M1:M4 | M2:M3 | M2:M4 | M3:M4 |
|---|---|---|---|---|---|---|---|---|---|---|---|
| g1 | 0.04 | 0.04 | 0.03 | 0.04 | g1 | 0.00 | 0.00 | 0.00 | 0.00 | 0.00 | 0.07 |
| g2 | 0.18 | 0.19 | 0.19 | 0.22 | g2 | 0.00 | 0.01 | 0.11 | 0.00 | 0.08 | 0.01 |
| g3 | 0.10 | 0.11 | 0.10 | 0.11 | g3 | 0.01 | 0.00 | 0.00 | 0.02 | 0.00 | 0.25 |
| g4 | 0.41 | 0.42 | 0.37 | 0.42 | g4 | 0.00 | 0.24 | 0.00 | 0.29 | 0.00 | 0.19 |
| g5 | 0.30 | 0.31 | 0.29 | 0.33 | g5 | 0.01 | 0.01 | 0.10 | 0.05 | 0.04 | 1.87 |
| sum | 1.04 | 1.07 | 0.97 | 1.11 | sum | 0.03 | 0.26 | 0.21 | 0.36 | 0.12 | 2.37 |

Another potentially robust way of estimating the parameters is to completely ignore the contribution of site patterns most associated with archaic/modern interbreeding (Waddell et al. 2011: 22, patterns NP, NH, NF, DP, and DNP) to the total fit. In doing this by simply failing to sum up the contribution of these cells, the restraint that the expected values sum to the observed values is also relaxed. Using the updated spreadsheet, this yield's $G^2$ = 1406.2, while g1-g5 are 0.0314, 0.1901, 0.0968, 0.3613 and 0.2844, respectively. These estimates are now very close to those at the $PD^\lambda$ optimum shown in table 2. They are also about 350 units better than in the previous calculations under the same conditions shown in Waddell et al. (2011, p22). The new calculations suggest that the approach used in Waddell et al. (2011), that is a locally corrected model by "recalibrating" with a very large simulated example, was valuable. Despite these issues, the corresponding parameters estimates in Waddell et al. (2011) are very very close to what we get here, and appear to be robust despite a worse overall fit. Table 2 also shows how the $PD^\lambda$ optimal model compares to some simulated results with an Approximate Bayesian Calculation (ABC) reticulate mixture model (Tan and Waddell, unpublished) of the four "species" trees shown in figure 9 of Waddell et al. (2011). That model allows Neanderthals to interbreed with the ancestors of all modern humans whose ancestors migrated out of Africa ~ 60 to 80 thousand years ago. The ABC method used a major simulation on the "Carter" super computer with randomly sampled parameters around their expected optimal values, with a large simulated sample being evaluated at each set of parameters and then used to estimate fit values. Then regressions of different orders (linear, quadratic, cubic and quartic) were fit with the parameters being the explanatory and fit being the dependent variable. A cubic fit was optimal by AIC with a quadratic fit being nearly as good. The parameter estimates from the quadratic regression were not quite as large as those obtained from the same samples by fitting quadratic curves to the marginal distributions of each parameter. While the marginal fits may be misleading due to correlated variables and a non-linear model, the full quadratic regression may suffer from having to estimate many cross terms, so it is not clear which are best.

The methods based on $PD^{\omega\lambda}$ fit even better with a minimum of around 1600. However, this is still not quite as good as the fit obtained with $G^2$ by ignoring the contribution of the five most obvious archaic/modern patterns. This fit is achieved at a very low value of $\lambda$ = -16.8 and P = 2.3 (which may be equated to a $P$ in section 3.3 of 2.3). The fit curves are slightly unusual with a slight inflection at larger values of $\lambda$ and a very flat trajectory at lower values of $\lambda$. Here, the moderately large positive α tells us to expect the biggest Power Divergence weighted residuals (either over or under) on the largest patterns. The parameter estimates from optimizing $PD^{\omega\lambda}$ are even larger than those of $PD^{\lambda}$ and from table 2 it can be seen they are the closest match to M2, the ABC marginal regression values. Whether these are even better estimates or if these parameter values are too high requires comparison to even better models. The best models accessible at present are those in table 8 of Waddell et al. (2011). Here the models which focus only on the informative sites (as here) achieve fits of around 1000 with $X^2$ (versus around $X^2 = 3000$ with the models here) and the parameter estimates are a best match to what is recovered with $PD^{\omega\lambda}$ (e.g., g1-g5, 0.04, 0.23, 0.13, 0.44, and 0.33 for the very best model in table 4, Waddell 2011:). This seems very encouraging. The lack of parabolic curve as seen with $PD^{\omega\lambda}$ in figure 3 is a feature also seen with polynomial powers to model errors in other contexts, such as distance models (e.g. Waddell and Azad 2009, where double minima, plateaus and points of inflexion may be seen).

It is also interesting to ask how many of the parameters g1-g5 are retained in the set that minimizes g%SD, g%AIC, etc. The results of leaving out each parameter in turn are shown in table 3. The results are ranked left to right by increasing g%BIC. Leaving out parameter g1, an edge in the species tree indicating Papuans and Han Chinese are closest relatives of the 7 genomes considered, increases g%BIC from 15.3 to 15.8%. Back transforming to a BIC value using equation 8 (that is, $N \times (2\ln[g\%BIC/100] + c)$) we find the difference is units is equal to $118 \times (2\ln(15.3/100) - 2\ln(15.8/100)) = 118 \times 2\ln(15.3/15.8) = 118 \times -0.0637 = -7.512$). How significant a decrease this is not immediately apparent. That is because the variance of the difference of two BIC values (which is the variance of the difference of the two $PD^{\omega\lambda}$ values) needs to be considered. Generally, for nested models the variance in the log likelihood often follows fairly closely that expected asymptotically under an adequate model. In that case -7.5 units difference is fairly significant, therefore we should favor the more highly parameterized model. If these were non-nested models, then the fluctuation of the difference (e.g., at $\lambda = 0$ or a log likelihood difference) can become a lot more extreme. An example is mentioned in Waddell et al. (2002) where the approximated difference of the variance of the log likelihood of two partly nested trees was over five times as big as expected in the former case.

The order in which the parameters are weighted by BIC are, in order of increasing importance, g1, α, g3, $\lambda$, g2, g5, g4 and finally, $\hat{\sigma}^2$. While the last value does not appear in table 3, its scale is readily estimated by noting that the sample "error" as measured by optimizing $PD^{\omega\lambda}$ is around 1600, when under the multinomial model it would be around 118. This means that $\hat{\sigma}^2$ is at least ~1600/118 or ~13.6 times as big as it would be in a standard multinomial likelihood calculation. This in turn would make all the differences in the log likelihood (or $PD^{\lambda}$ value) around $(1600/118)^{0.5}$ bigger, and the differences in BIC, AIC, etc. values ~1600/118 times bigger, which is a major impact.

Figure 4 tracks the changes in parameter values with respect to g%BIC as other parameters are included or excluded. In figure 4(b), the changes are tracked relative to their size on the optimal model. One important use of figure 4(a) might be to guestimate where the parameter value might end up at under the true model, which by definition, resides at g% = 0. The curves generated in these Excel charts are Bezier curves and the overall trend, particularly for the better fitting models might be an indication of how stable a parameter is over a range of models.

It is interesting to use figure 4 to look at where a parameter estimate is most unexpected. For example, for parameter α notice the major blip in the curve at around g%BIC = 21. This is caused when parameter $\lambda$ is set to zero (reverting back to a traditional maximum likelihood model). Loss of parameter $\lambda$ induces a noticeable irregularity not only in the estimate for parameter α, but also in every other parameter estimate, all towards markedly smaller absolute values. In contrast, a smaller blip in λ occurs when parameter α is excluded from the model. It should be noted, that this guestimate of the stability of a parameter to the overall structure of the model is very much predicated or conditional on the other parameters in the model. In this regard, another notable blip is seen in the large increase in g3 when g4 is excluded (yet other parameter estimates continue to follow their local trends). Similarly, g1 and g3 are markedly affected when parameters g2 is excluded. These pairs of parameters seem to be negatively correlated with each other, so one increases markedly in its estimated magnitude when the other disappears.

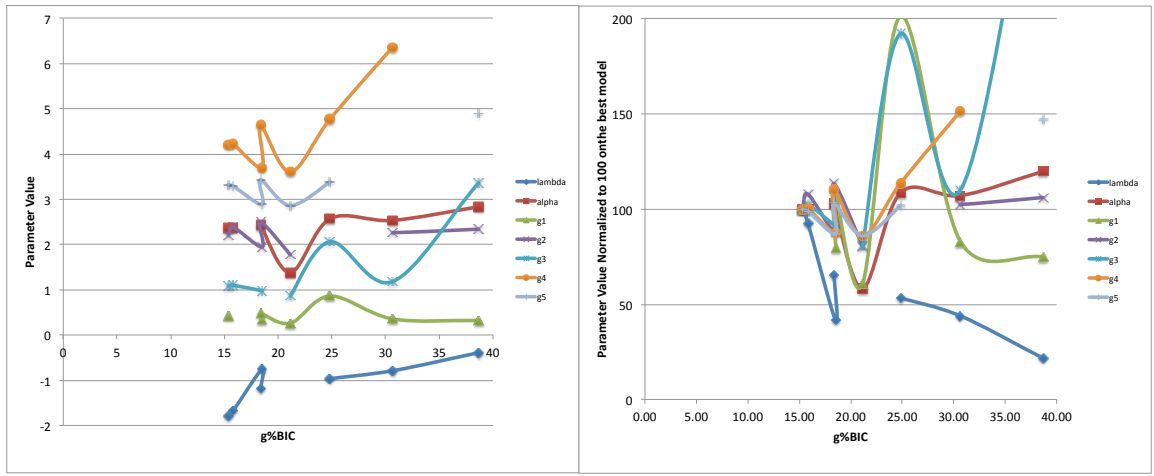

Figure 4. Parameter values plotted against fit measured by g%BIC_$PD^{\alpha\lambda}$ with all parameters and with each parameter left out of the model (to help guestimate robustness to latent variables; actual values shown in table 3). The values for lambda are divided by 10, while those for g1-g5 are multiplied by 10 to assist in plotting. The plot on the right is the same as that on the left, except all parameters are normalized to being 100 on the best model. A missing point in a line occurs when that particular parameter is excluded from the model.

Table 3. Parameter estimates for model fitted to the Neanderthal data by $PD^{\alpha\lambda}$ as each variable, except the estimated variance, is removed. The first model has 8 freely optimized parameters while all the others have 7. The asterisk "*" indicates that parameter is missing from the submodels considered.

| PD   | 1597   | 1777   | 2351   | 2230  | 2832  | 4426  | 7034  | 10229 |
|------|--------|--------|--------|-------|-------|-------|-------|-------|
| g%   | 13.3   | 14.0   | 16.3   | 16.4  | 18.7  | 22.0  | 27.2  | 34.3  |
| AIC  | 14.0   | 14.6   | 16.9   | 17.0  | 19.5  | 22.9  | 28.2  | 35.6  |
| AICc | 14.0   | 14.6   | 17.0   | 17.1  | 19.5  | 23.0  | 28.3  | 35.7  |
| BIC  | 15.3   | 15.8   | 18.4   | 18.5  | 21.1  | 24.8  | 30.7  | 38.7  |
| BICc | 15.5   | 16.0   | 18.5   | 18.7  | 21.3  | 25.1  | 30.9  | 39.0  |
| λ    | -18.01 | -16.68 | -11.77 | -7.51 | *     | -9.60 | -7.91 | -3.92 |
| α    | 2.365  | 2.358  | 2.428  | *     | 1.381 | 2.569 | 2.533 | 2.833 |
| g1   | 0.043  | *      | 0.048  | 0.034 | 0.026 | 0.086 | 0.035 | 0.032 |
| g2   | 0.220  | 0.238  | 0.250  | 0.194 | 0.178 | *     | 0.225 | 0.234 |
| g3   | 0.107  | 0.110  | *      | 0.098 | 0.087 | 0.206 | 0.118 | 0.336 |
| g4   | 0.420  | 0.422  | 0.464  | 0.370 | 0.360 | 0.477 | 0.636 | *     |
| g5   | 0.332  | 0.329  | 0.343  | 0.289 | 0.286 | 0.339 | *     | 0.489 |

In terms of apparent stability of parameter estimates in terms of proportional changes, it would seem that figure 4(b) suggests that parameters g5, g4 and g2 show moderate stability when

g%BIC is less than 25. In contrast, the other parameters are still apparently unpredictable and we expect that parameter $\hat{\sigma}^2$ will reduce considerably as better models are found. Indeed, in Waddell et al. (2012) there is the prediction for this data that $PD^{\lambda=1}$ (a minimum $X^2$ estimate) should eventually reach a value of around 160 for a truly reasonable model (excluding intra chromosomal events, such as linkage, chromosome wide selection or mutation bias). As mentioned earlier this is a g%SD of around 4-5% for this much data. If the trend in figure 4(b) is not obvious in at least the better models, and sufficiently extended to give some confidence of extrapolating to g%0, it would seem very difficult to guestimate from data and model alone the range of values we expect each parameter to achieve in the future as models improve.

It is useful to emphasize that the afore mentioned assessment of model stability is not that of simply the residual error, which can perhaps best be measured by residual resampling in this context (Waddell and Azad 2009). Rather it is a broad assessment of what might happen to all parameter estimates as latent (hidden) variables (parameters) are incorporated into the model. This is potentially a much bigger problem in statistical modeling of genetic data than is commonly realized, especially when we are dealing with g% errors 3-4 times as big as they are expected to be, as we are in this case.

The actual optimizations in table 3 were performed to minimize g%AIC. A useful property of all the g% measures is that for the same $k$, they are always monotonic functions of each other, thus only one needs to be minimized if searching across $k$. They are very similar to, but slightly different from minimizing $PD^\lambda$. The reason for this is that they incorporate a correction term for the expected variance when it is modeled as a power function of the expected value with parameter α. As seen in equation 1, this requires addition of an extra normalization term other than the $2/(\lambda(\lambda+1))$ normalization term in $PD^\lambda$ itself.

## 3.9 Site patterns and more robust modeling

In this section the main purpose is to look at modeling the Neanderthal data with two factors revealed above. One of these is the use the $PD^\lambda$ and $PD^{\alpha\lambda}$ families. The second feature is to combine this with censoring the five site patterns known to be most influenced by Denisovan's and Neanderthals breeding with modern humans (patterns NP ,NH, NF, DP, DNP). Upon completion we are also interested to look at the residuals of the three models of maximum likelihood, minimum $PD^\lambda$ and minimum $PD^{\alpha\lambda}$.

The results of estimating parameters with ML, minimum $PD^\lambda$ and minimum $PD^{\alpha\lambda}$ are shown in table 4. The parameters of the minimum $PD^\lambda$ model are now very close to those of the ML model, although the parameter λ = 3.1 is a little extreme and now penalizes inliers more than outliers. The change in fit with respect to λ is much less than it had been, which is consistent an expectation under the multinomial model. The minimum $PD^{\alpha\lambda}$ still model shows a major improvement in fit and it wants to avoid punishing the unexpectedly large mostly positive deviations on the larger cells, hence α = 2.1 (P equivalent = 2.1) and λ = -7.2.

Table 5 shows the largest residuals for each model. Two patterns of particular interest are NSYFHP and DSYFHP. The frequency of these patterns are one line of evidence to probe the hard to test hypothesis that either Neanderthals or Denisovan's interbreed with earlier humans such as *Homo erectus* (Krause et al. 2010, Waddell et al. 2011). Two possible causes upsetting these patterns expected frequencies are that early Neanderthal's (or late European *Homo heidelbergensis*) introgressed with the ancestors of all modern humans in Africa perhaps 150 to 250 thousand years ago. Another, is that the Denisova individual is a hybrid of mostly Asian *Homo heidelbergensis* plus a fraction (perhaps 5 to 20 percent) of earlier Asian hominids, of which *Homo erectus* (e.g., Peking man) is the only one well characterized. Analyses summarized in Waddell et al. (2011) suggest the later seems more likely.

Table 4. Fit and parameter estimates of maximum likelihood (= minimum $G^2$), minimum $PD^\lambda$ and minimum $PD^{\alpha\lambda}$.

| k | 6 | 7 | 8 |
|---|---|---|---|
| PD | 1406.2 | 1384.5 | 1002.4 |
| g% | 13.1 | 13.0 | 10.7 |
| AIC | 13.5 | 13.5 | 11.2 |
| AICc | 13.5 | 13.6 | 11.3 |
| BIC | 14.5 | 14.7 | 12.3 |
| BICc | 14.6 | 14.8 | 12.5 |
| $\lambda$ | 0 | 3.1 | -7.2 |
| $\alpha$ | 1 | 1 | 2.1 |
| g1 | 0.03 | 0.03 | 0.04 |
| g2 | 0.19 | 0.19 | 0.20 |
| g3 | 0.10 | 0.10 | 0.10 |
| g4 | 0.36 | 0.36 | 0.41 |
| g5 | 0.28 | 0.28 | 0.32 |

The apparent lack of Denisovan alleles on the X chromosome suggested that some of these archaic interbreeding events were male biased, that is archaic males mating with modern females (Waddell, 2011). This was formerly dubbed the "archaic Ron Jeremy" hypothesis, after the well-known American thespian. Formerly known, because a journal editor has recently urged us to alter our manuscript, to avoid confusion with a "Ron Jeremy Event", which they referenced to the Urban Dictionary. The new synonymy is the "lecherous archaic man" hypothesis. Looking at the residuals, evidence for introgression of *Homo erectus* into the Denisovan is the marked deficiency of the DSYFHP allele in contrast to a slight excess of the NSYFHP allele. The other site patterns most out of balance in the ML and minimum $PD^\lambda$ models are the "species tree" patterns DN and FHP. The deficiency of DN is consistent with Denisovan-*Homo erectus* interbreeding. It is unclear what might have caused FHP, the most prominent "out of Africa" allele in this data, to be in excess.

Table 5. All cells with misfits of 60 or more for the ML, minimum $PD^{\lambda=3.1}$ and minimum $PD^{\alpha=2.1,\lambda=-7.2}$ models, all ignoring the contribution of the cells NP, NH, NF, DP, and DNP.

| Patt | Obs | Exp | Res | $G^2$ | Pat | Obs | Exp | Res | $PD^\lambda$ | Patt | Obs | Exp | Res | $PD^{\alpha\lambda}$ |
|---|---|---|---|---|---|---|---|---|---|---|---|---|---|---|
| DN | 11849 | 12980 | -1131 | 101.5 | FHP | 5340 | 4652 | 688 | 112.3 | DSHP | 218 | 268 | -50 | 50.2 |
| FHP | 5340 | 4681 | 659 | 88.7 | DN | 11849 | 12945 | -1096 | 87.5 | DNSYFH | 1179 | 865 | 314 | 43.5 |
| DSYFHP | 4069 | 4689 | -620 | 85.9 | DSYFHP | 4069 | 4672 | -603 | 71.2 | NSYFH | 502 | 376 | 126 | 42.0 |
| DNSYFH | 1179 | 947 | 232 | 52.6 | SYFH | 1643 | 1369 | 274 | 63.0 | NSFH | 325 | 246 | 79 | 41.6 |
| SYFH | 1643 | 1368 | 275 | 51.7 | DNSYFH | 1179 | 955 | 224 | 61.4 | NFH | 429 | 325 | 104 | 40.1 |
| YF | 3141 | 2766 | 375 | 48.7 | YF | 3141 | 2763 | 378 | 56.9 | NSFP | 204 | 246 | -42 | 39.1 |
| SF | 2432 | 2114 | 316 | 45.5 | SF | 2432 | 2115 | 317 | 52.5 | SYP | 1014 | 1223 | -209 | 34.3 |

Fitting the model with minimum $PD^{\alpha\lambda}$, minus the effect of the five alleles most in excess due to archaic interbreeding, yields a model with parameters that are an even closer match to those of the ABC models, in particular due to parameters g2 and g5 decreasing. The most prominent residuals are shown in table 5. Two of them may be interpretable. The pattern NFH is the most prominent of the next most common patterns expected due to Neanderthals interbreeding with the ancestors of the out-of-Africa people. This is test able in future if it goes away by modeling this effect (as was done in Waddell et al. 2011). The marked excess of pattern DNSYFH, can also be interpreted as an excess of its compliment, Chimp (or outgroup) with Papuan. This is consistent with Papuan unique amongst the modern humans sampled, holding even more archaic pre *Homo heidelbergensis* alleles. This is consistent with evidence of *Homo erectus* and the even more primitive *Homo floresiensis* occupying the South East Asia region when modern humans occupied the region, perhaps 50 to 70 thousand years ago. While the fit of the pattern DSYFHP is improved to 17.5 due to an effective polynomial power of around 3.1, its residual for the model $PD^{\omega\lambda}$ has actually become more extreme going to -654 from -603. At the same time the residual for NSYFHP has decreased from 104 to 53. Thus, the evidence for a Denisovan-*Homo erectus*-like interbreeding event remains stable.

## 4 Discussion

It is useful to note that after fitting the minimum $PD^{\omega\lambda}$ model there is a weighted "residual" for each cell that can be renormalized, randomly sampled, reallocated and rescaled before adding back to the expected data vector. Alternatively, it is possible simulate data that come from a distribution with properties consistent with the observed α and λ. Such residual resampling methods have been found to be particularly appropriate and robust with phylogenetic models based on distances (Waddell and Azad 2009).

In terms of entropy (the term used most often in information theory), using the statistic $PD^{\omega\lambda}$ we have learned how the disorder (uncertainty or residuals) seems to arise in order to find a most useful measure of entropy. This guide is in turn used to minimize the entropy between model and the data and so hopefully maximize the useful and usable information. Another way of saying this is that we have sought the fit function that is least influenced by inliers and outliers (subject to minimizing the needed number of parameters, so that we can expect to minimize the residuals on a truly new sample of data). The term truly new is important, since in some systems such as a genome, it is very unclear what a truly new sample of data is, as so many of the generating parameters (e.g., base composition bias, equals the frequency of the nucleotides A, C, G and T) may be biased throughout the whole genome. Thus, the answer cannot automatically be more data from the same source to produce a truly independent sample.

It is useful to consider further what each of the g% measures may mean. The g%SD measure is a standard unbiased estimator of the deviation of the current data from the model. Deviation is measured as a normalized weighted root mean square error (expressed as a percentage). This may then be directly interpreted by assuming that all error about the true model is stochastic (according to the estimation methods assumptions) and centered on the assumed tree. The g%AICc statistic seems to be a reduced bias estimator (compared to g%AIC, which is asymptotically justified) estimator of how much normalized weighted root mean square error (expressed as a percentage) to expect in fitting new observations to the model based on essentially the same assumptions as g%SD. BICc is conjectured to be a reduced bias estimator (compared to g%BIC) of how much normalized weighted root mean square error (expressed as a percentage) to expect if fitting new observations to the current model weighted so as to converge to the true model without spurious parameters (and assuming uniform priors on all models, and that there is an accessible true model).

To some extent the easiest interpretation of the g% measures is predicated on the expectation that the total error represents sampling error. Often, the total residual error it will be dramatically inflated by systematic error. It is almost certain that tree search and some "non-identifiability" of errors will decrease the total error from the true model (e.g., weighted tree). The effect of tree search can be given a loose lower bound by residual resampling to create "replicate" data sets, followed by tree search and looking at how much the sum of squares shrinks during the search (Waddell and Azad 2009). The "non-identifiability" of errors is more difficult. Trees "eat" certain types of error, particularly for distances from sequences. That is, if the data evolved on a tree, then some errors on distances are highly correlated, as they represent a single source of error on an edge of the tree. Other effects of both long edges attract and long edges repel (Waddell 1995) can cause any replicate distance matrix under the true model to be most additive on an incorrect tree, often with widely erroneous edge lengths. Thus, the total errors being estimated by the current g% measure are biased downwards and are predicated on the reconstructed edge lengths being unbiased estimators of the true edge lengths and the data being most additive on the generating tree. Given that all the biases here tend to potentially strongly reduce the total residual, then g% measures should be taken as a conservative warning of the total error to be expected.

Putting it another way, the total sum of squares error of data to a tree model is a mixture of stochastic error (more equals less precision) and bias (more equals less accuracy). Here, the total error is not being partitioned, but it is all assumed to be stochastic error and hence decreases the expected precision. When no further attempt is made to separate these two components apart, a sensible forecasting strategy for the whole model is to group the whole lot together and treat it as stochastic error and use it to gauge the precision of parameters and predicting new data. Thus, g% measures are a type of quasi-accuracy. In a sociological context, this is also an appropriately strong impetus for researchers to attempt to reduce the total error and hence improve the reportable quasi-accuracy of their methods. At present in the science of phylopop, and to some extent genomics, it is often the mistaken reporting of high estimates of precision from an i.i.d. bootstrap as accuracy, that is a major impediment to better science.

Future directions include implementing more diverse functions, such as an exponential to model how the variance of distances or discrete cell counts change with the size of the cell (Waddell et al. 2010c), in order to maximize the information content by minimizing measures such as g%AIC. In terms of modeling of the Denisova/Neanderthal site pattern data, the next big step is to model reticulate coalescent events such as those examined in Waddell et al. (2011) with the new criterion of minimum $PD^{\omega\lambda}$ modeling.

## Acknowledgements

This work was supported by National Institutes of Health grant 5R01LM008626 and the Recovery Act Supplement 3R01LM008626-05S1 (to PJW). Thanks to Kenneth Burnham, Hiro Kishino and Mike Steel for helpful discussions, plus Jorge Ramos for corrections to the spreadsheet used in Waddell et al. (2011). Thanks to David Bryant for his hospitality, discussions, and editorial comments. Thanks to Steve Kelly and other staff of itap at Purdue for assistance running the new super computer "Carter."

## Author contributions

PJW originated the research, developed methods, gathered data, ran analyses, interpreted analyses, prepared figures and wrote the manuscript. XT, implemented methods in PERL, ran analyses and commented on the manuscript. PJW and XT derived equations.